\documentclass[superscriptaddress,aps,letterpaper,10pt,twocolumn,floats,showpacs,amsmath,amsfonts,amssymb,prl]{revtex4-1}

\usepackage{graphicx}
\usepackage[colorlinks=true,citecolor=blue]{hyperref}
\usepackage[caption=false]{subfig}
\usepackage{pstricks}
\usepackage{pst-node}
\usepackage{amsmath}
\usepackage{amsbsy}
\usepackage{verbatim}
\begin{document}


\title{Thermodynamics of Quantum Information Flows}

\author{Krzysztof Ptaszy\'{n}ski}
\affiliation{Institute of Molecular Physics, Polish Academy of Sciences, ul.~M.~Smoluchowskiego 17, 60-179 Pozna\'{n}, Poland}
\email{krzysztof.ptaszynski@ifmpan.poznan.pl}
\author{Massimiliano Esposito}
\affiliation{Complex Systems and Statistical Mechanics, Physics and Materials Science Research Unit, University of Luxembourg, L-1511 Luxembourg, Luxembourg}

\date{\today}

\begin{abstract}
We report two results complementing the second law of thermodynamics for Markovian open quantum systems coupled to multiple reservoirs with different temperatures and chemical potentials. First, we derive a nonequilibrium free energy inequality providing an upper bound for a maximum power output, which for systems with inhomogeneous temperature is not equivalent to the Clausius inequality. Secondly, we derive local Clausius and free energy inequalities for subsystems of a composite system. These inequalities differ from the total system one by the presence of an information-related contribution and build the ground for thermodynamics of quantum information processing. Our theory is used to study an autonomous Maxwell demon.
\end{abstract}

\maketitle


The second law of thermodynamics is one of the main principles of physics. Within  equilibrium thermodynamics there exist two equivalent formulations of this law. The first, referred to as the \textit{Clausius inequality}, states that the sum of the entropy change of the system $\Delta S$ and the entropy exchanged with the environment $\Delta S_\text{env}$ during the transition between two equilibrium states is nonnegative: ${\Delta S+ \Delta S_\text{env} \geq 0}$. The exchanged entropy can be further expressed as $\Delta S_\text{env}=-Q/T$, where $Q$ is the heat delivered to the system. An alternative formulation, referred to as the \textit{free energy inequality}, states that during the transition between two equilibrium states $W-\Delta F \geq 0$, where $W$ is the work performed on the system and $F=E-TS$ is the free energy (here $E$ denotes the internal energy). The latter formulation can be obtained from the former by using the first law of thermodynamics $\Delta E=W+Q$.

Whereas these standard definitions of the second law apply when considering transitions between equilibrium states, the last few decades have brought significant progress towards generalizing them to both classical~\cite{evans1993, gallavotti1995, lebowitz1999, jarzynski1999, seifert2005, esposito2010, seifert2012} and quantum~\cite{spohn1978, esposito2010b, cuetara2016} systems far from equilibrium. The most common formulation generalizes the Clausius inequality by stating that the average entropy production $\sigma$ is nonnegative. For a large class of systems~\cite{jarzynski1999, esposito2010b, esposito2010, seifert2005} the entropy production can be defined as ${\sigma \equiv \Delta S-\sum_\alpha Q_\alpha \beta_\alpha}$, where $\Delta S$ is the change of the Shannon or the von Neumann entropy of the system (which is well defined also out of equilibrium) and $Q_\alpha$ is the heat delivered to the system from the reservoir $\alpha$ with the inverse temperature $\beta_\alpha$; additionally, in Markovian systems the entropy production rate $\dot{\sigma}$ is always nonnegative~\cite{seifert2012, cuetara2016}. Formulations generalizing the free energy inequality~\cite{esposito2011, parrondo2015, strasberg2017, miyahara2018} are much less common and have been so far confined mainly to systems coupled to an environment with a homogeneous temperature; for an exception, see Ref.~\cite{miyahara2018}.

These developments have also brought a deeper understanding of the relation between thermodynamics and the information theory~\cite{parrondo2015, goold2016}. One of the most important achievements is related to the field of thermodynamics of feedback-controlled systems~\cite{maruyama2009}. Following the groundbreaking ideas of Maxwell demon~\cite{maxwell1995} and Szilard engine~\cite{szilard1929}, it was verified both theoretically~\cite{lloyd1997, kim2011, schaller2011, mandal2012, park2013} and experimentally~\cite{koski2014, koski2015, koski2016, camati2016, cottet2017, chida2017} that by employing feedback one can reduce entropy of the system without exchanging heat. In such a case modified Clausius inequalities, which relate the entropy change to the information flow, have to be applied~\cite{sagawa2010, esposito2011, parrondo2015, sagawa2013, horowitz2010, morikuni2011, abreu2012, deffner2013, strasberg2014, barato2014, strasberg2017, chapman2015, hartich2014, horowitz2014, horowitz2014b, horowitz2015, miyahara2018}. It was also realized that the feedback control does not require the presence of any intelligent being (as in the original idea of Maxwell), but may be performed by an autonomous stochastic system coupled to the controlled one~\cite{strasberg2013}. A consistent mathematical description of thermodynamics of autonomous information flow has been, however, so far confined mainly to classical stochastic systems with a special topology of network of jump processes, referred to as bipartite~\cite{hartich2014, horowitz2014} or, in general, multipartite~\cite{horowitz2015} systems.

Our work adds two new contributions to the field. First, we generalize the free energy inequality to Markovian open quantum systems coupled to reservoirs with different temperatures and show that this formulation of the second law is in general not equivalent to the Clausius inequality. Secondly, we formulate a consistent thermodynamic formalism describing thermodynamics of continuous information flow in a generic composite open quantum system and demonstrate the relation between the information and the nonequilibrium free energy. The applicability of our results is demonstrated on a quantum autonomous Maxwell demon based on quantum dots.

\textit{Nonequilibrium Clausius inequality}. We consider a generic open quantum system weakly coupled to $N$ equilibrated reservoirs $\alpha$ with temperatures $T_\alpha$ (inverse temperatures ${\beta_\alpha \equiv 1/T_\alpha}$) and chemical potentials $\mu_\alpha$, described by the time-independent Hamiltonian
\begin{align}
\hat{H}=\hat{H}_S+\hat{H}_B+\hat{H}_{\text{I}},
\end{align}
where $\hat{H}_S$, $\hat{H}_B$, $\hat{H}_{\text{I}}$ are, correspondingly, the Hamiltonian of the system, reservoirs and interaction of the system with the reservoirs. Within the Markov approximation the (reduced) density matrix of the system evolves according to the master equation~\cite{breuer2002}
\begin{align}
d_t \rho=- i [\hat{H}_{\text{eff}}, \rho] + \mathcal{D}\rho,
\end{align}
where $\rho$ is the density matrix, $d_t$ denotes the total derivative of the function, $\hat{H}_{\text{eff}}$ is the effective Hamiltonian of the system (it may differ from $\hat{H}_S$ due to coupling to the environment~\cite{breuer2002}), and $\mathcal{D}$ is the superoperator describing the dissipative dynamics. Here and from here on we take $\hbar=k_B=1$. We further assume that the dissipator $\mathcal{D}$ is of Lindblad form, thus ensuring a completely-positive trace-preserving dynamics~\cite{lindblad1976, gorini1976}, and that $\hat{H}_{\text{eff}}$ commutes with $\hat{H}_S$, which is justified by the perturbation theory (cf., the Supplementary Material~\cite{supp}). Furthermore, within the Markov approximation the dissipation is additive, i.e., the superoperator $\mathcal{D}$ can be represented as a sum of dissipators associated with each reservoir, denoted as $\mathcal{D}^\alpha$: $\mathcal{D}=\sum_\alpha \mathcal{D}^\alpha$~\cite{schaller2014, cuetara2016}. For violation of additivity beyond the weak coupling regime, see Refs.~\cite{chan2014, giusteri2017, mitchison2018, friedman2018}. 

We also assume, that the grand canonical equilibrium state (Gibbs state) with respect to the reservoir $\alpha$
\begin{align}
\rho_\text{eq}^\alpha = Z_{\beta_\alpha,\mu_\alpha}^{-1} e^{-\beta_\alpha \left(\hat{H}_S-\mu_\alpha \hat{N} \right)},
\end{align}
is a stationary state of $\mathcal{D}^\alpha$, i.e., ${\mathcal{D}^\alpha \rho_\text{eq}^\alpha=0}$~\cite{cuetara2016}; here $Z_{\beta_\alpha,\mu_\alpha}=\text{Tr} \{\exp[ -\beta_\alpha (\hat{H}_S-\mu_\alpha \hat{N})]\}$ is the partition function and $\hat{N}$ is the particle number operator. This assumption guarantees that for an arbitrary form of dissipator the Gibbs state is a stationary state at equilibrium (i.e., for equal temperatures and chemical potentials of the reservoirs), which is true for systems weakly coupled to the environment~\cite{breuer2002}. Let us then apply the Spohn's inequality~\cite{spohn1978}
\begin{align}
-\text{Tr} \left[ \left( \mathcal{D}^\alpha \rho \right) \left( \ln \rho-\ln \rho_\text{eq}^\alpha \right) \right] \geq 0,
\end{align}
which is valid for any superoperator $\mathcal{D}^\alpha$ of Lindblad form with a steady state $\rho_\text{eq}^\alpha$ (not necessarily a unique steady state). As a result, one obtains the \textit{partial Clausius inequality} for entropy production associated with each dissipator~\cite{cuetara2016}
\begin{align} \label{part2ndlaw}
\dot{\sigma}_\alpha= \dot{S}^\alpha-\beta_\alpha \dot{Q}_\alpha \geq 0,
\end{align}
where 
\begin{align}
\dot{S}^\alpha=-\text{Tr} \left[ \left( \mathcal{D}^\alpha \rho \right) \ln \rho \right],
\end{align}
is the rate of change of the von Neumann entropy of the system $S=-\text{Tr} (\rho \ln \rho)$ due to the dissipator $\mathcal{D}^\alpha$ and 
\begin{align}
\dot{Q}_\alpha=\text{Tr} \left[ \left( \mathcal{D}^\alpha \rho \right) \left( \hat{H}_S-\mu_\alpha \hat{N} \right) \right],
\end{align}
is the heat current from the reservoir $\alpha$. Summing all the rates $\dot{S}^\alpha$ one gets the total derivative of the von Neumann entropy: $d_t S=\sum_\alpha \dot{S}^\alpha$. Therefore, summing up Eq.~\eqref{part2ndlaw} over the reservoirs $\alpha$ one recovers the standard Clausius inequality~\cite{cuetara2016}
\begin{align} \label{tot2ndlaw}
\dot{\sigma} \equiv \sum_{\alpha} \dot{\sigma}_\alpha=d_t S-\sum_{\alpha } \beta_\alpha \dot{Q}_\alpha \geq 0,
\end{align}
where $\dot{\sigma}$ is the total entropy production rate. We note that the rates $\dot{S}^\alpha$ can be non-zero also at the steady state, when $d_t S=0$ and the total entropy production is fully determined by the heat flows.

\textit{Nonequilibrium free energy inequality.} Let us now define energy and work currents to the lead $\alpha$ as
\begin{align}
\dot{E}_\alpha &=\text{Tr} \left[ \left( \mathcal{D}^\alpha \rho \right) \hat{H}_S  \right], \\
\dot{W}_\alpha &=\mu_\alpha \text{Tr} \left[ \left( \mathcal{D}^\alpha \rho \right) \hat{N} \right],
\end{align}
such that $\dot{E}_\alpha=\dot{Q}_\alpha+\dot{W}_\alpha$ and $\sum_{\alpha} \dot{E}_\alpha=d_tE$, where $E=\text{Tr} (\rho \hat{H}_S)$ is the internal energy. Since we assume the Hamiltonian to be time independent, we consider only chemical and not mechanical work. Multiplying Eq.~\eqref{part2ndlaw} by $T_\alpha$ and replacing $\dot{Q}_\alpha \rightarrow \dot{E}_\alpha-\dot{W}_\alpha$ one gets
\begin{align} \label{partfreeenineq}
T_\alpha \dot{\sigma}_\alpha=\dot{W}_\alpha-\dot{\mathcal{F}}_\alpha \geq 0,
\end{align}
where $\dot{\mathcal{F}}_\alpha \equiv \dot{E}_\alpha-T_\alpha \dot{S}^\alpha$ is the \textit{partial nonequilibrium free energy rate}. Summing over $\alpha$ one obtains the \textit{nonequilibrium free energy inequality}
\begin{align} \label{freeenineq}
\sum_{\alpha} T_\alpha \dot{\sigma}_\alpha=\dot{W}-\dot{\mathcal{F}} \geq 0,
\end{align}
where $\dot{W} \equiv \sum_\alpha \dot{W}_\alpha$ is the total work rate and
\begin{align}
\dot{\mathcal{F}} \equiv \sum_\alpha \dot{\mathcal{F}}_\alpha = d_t E-\sum_\alpha T_\alpha \dot{S}^\alpha,
\end{align}
is the \textit{total nonequilibrium free energy rate}. Equation~\eqref{freeenineq} is a complementary formulation of the second law of thermodynamics. From a practical point of view, it provides an upper bound for the maximum work output. At the steady state $d_t E=0$ and thus the system can perform work ($\dot{W}<0$) only when a temperature difference between the reservoirs is present.

Let us emphasize, that Eqs.~\eqref{tot2ndlaw} and~\eqref{freeenineq} are in general not equivalent; the former corresponds to the sum of partial Clausius inequalities [Eq.~\eqref{part2ndlaw}], whereas the latter to the weighted sum, in which Eq.~\eqref{part2ndlaw} is multiplied by a local temperature $T_\alpha$. They become equivalent only when the system is attached to an isothermal environment, i.e., {$T_\alpha=T$}. Then the rate $\dot{\mathcal{F}}$ can be identified as the total derivative of the state function $F$: $\dot{\mathcal{F}}= d_t F = d_t (E-TS)$. At the steady state $d_t F=0$ and thus $\dot{W}>0$. This corresponds to the \textit{Kelvin-Planck statement of the second law}, according to which one cannot continuously generate work by cooling an isothermal environment.

\textit{Local Clausius inequality}. Let us now consider a system made of two coupled subsystems described by the Hamiltonian
\begin{align} \label{ham2comp}
\hat{H}_S=\hat{H}_1 + \hat{H}_2 +\hat{H}_{12},
\end{align}
where $\hat{H}_i$ is the Hamiltonian of the subsystem $i=1,2$ and $\hat{H}_{12}$ is the interaction Hamiltonian. We also assume that each subsystem is attached to a separate set of reservoirs; baths coupled with the subsystem $i$ will be then denoted as $\alpha_i$. By summing Eq.~\eqref{part2ndlaw} over the reservoirs $\alpha_i$ one obtains
\begin{align} \label{partsummed}
\dot{\sigma}_i\equiv\sum_{\alpha_i} \dot{\sigma}_{\alpha_i}=\sum_{\alpha_i } \dot{S}^{\alpha_i}-\sum_{\alpha_i } \beta_{\alpha_i} \dot{Q}_{\alpha_i} \geq 0.
\end{align}
Here $\dot{\sigma}_i=\sum_{\alpha_i} \dot{\sigma}_{\alpha_i}$ denotes the local entropy production; it is an extensive quantity, i.e., $\dot{\sigma}=\dot{\sigma}_1+\dot{\sigma}_2$. We will now transform Eq.~\eqref{partsummed} to a form illustrating the relation between entropy and information. Let us remind that the quantum mutual information is defined as $I_{12}=S_1+S_2-S$ where $S_i=-\text{Tr} (\rho_i \ln \rho_i)$ is the von Neumann entropy of the subsystem $i$~\cite{wilde2013} (here $\rho_i$ is the reduced density matrix of the subsystem $i$). We can then separate the total derivative of the mutual information into two contributions: $d_t I_{12}=\dot{I}_1+\dot{I}_2$ where
\begin{align}
\dot{I}_i \equiv d_t S_i-\sum_{\alpha_i} \dot{S}^{\alpha_i}.
\end{align} 
Here we have applied the identity $\sum_\alpha \dot{S}^\alpha=d_t S$. The rate $\dot{I}_i$ can be calculated as
\begin{align}
\dot{I}_i =-\text{Tr} \left( d_t \rho_i\ln \rho_i \right)+\text{Tr} \left[ \left( \mathcal{D}_i \rho \right) \ln \rho \right],
\end{align}
where ${\mathcal{D}_i=\sum_{\alpha_i } \mathcal{D}^{\alpha_i}}$ is the dissipator associated with the subsystem $i$. The rate $\dot{I}_i$ can be further decomposed into contributions related to the unitary and the dissipative dynamics; see the Supplementary Material~\cite{supp} for details.

Replacing $\sum_{\alpha_i} \dot{S}^{\alpha_i} \rightarrow d_t S_i-\dot{I}_i$ in Eq.~\eqref{partsummed} one obtains the \textit{local Clausius inequlity} relating the entropy balance of the subsystem $i$ to the information flow:
\begin{align} \label{2ndlawloc}
\dot{\sigma}_i=d_t S_i-\sum_{\alpha_i } \beta_{\alpha_i} \dot{Q}_{\alpha_i} -\dot{I}_i \geq 0.
\end{align}
This inequality is identical in form to the one previously derived in Ref.~\cite{horowitz2014}. However, our result is much more general. First, it enables to describe systems undergoing a quantum dynamics formulated in terms of a density matrix, whereas the former approach was purely classical and formulated in terms of probabilities. Second, our result has a much wider range of applicability even in the classical limit. Indeed, the approach from Ref.~\cite{horowitz2014} was restricted to so-called ``bipartite'' systems, which exclude stochastic transitions generating a simultaneous change of states of both subsystems. However, two-component open quantum systems are in general not bipartite even when their populations obey a classical master equation. Instead, our only requirement is that the dissipation is additive, i.e., one can split the dissipator $\mathcal{D}$ into contributions $\mathcal{D}_i$ in a physically meaningful way. The system can become bipartite when the total Hamiltonian $\hat{H}_S$ commutes with the subsystem Hamiltonian $\hat{H}_i$ and one applies the effectively classical description by means of the secular (rotating wave) approximation; in such a case our approach reduces to that from Ref.~\cite{horowitz2014}. We discuss these issues is detail in the Supplementary Material~\cite{supp}.

Let us finally emphasize, that all the previous discussion can be easily generalized to the multicomponent systems consisting of $M$ subsystems. Then $\sum_i \dot{I}_i=d_t I_{1,...,M}$, where $I_{1,...,M}=\sum_i S_i-S$ is the multipartite mutual information~\cite{watanabe1960}.

\textit{Local free energy inequality}. Analogously, summing up Eq.~\eqref{partfreeenineq} over reservoirs $\alpha_i$ we derive an inequality describing the nonequilibrium free energy balance for a single subsystem:
\begin{align} \label{locfreeenineq}
\sum_{\alpha_i} T_{\alpha_i} \dot{\sigma}_{\alpha_i}=\dot{W}_i-\dot{\mathcal{F}}_i \geq 0,
\end{align}
where $\dot{W}_i \equiv \sum_{\alpha_i} \dot{W}_{\alpha_i}$ and $\dot{\mathcal{F}}_i \equiv \sum_{\alpha_i} \dot{\mathcal{F}}_{\alpha_i}$. 

As in the case of the total system, Eqs.~\eqref{2ndlawloc} and~\eqref{locfreeenineq} are nonequivalent and complementary formulations of the local second law of thermodynamics. They become equivalent when the subsystem $i$ is coupled to an isothermal environment with a single temperature $T_i$. Then $T_i \dot{I}_i=-T_i \sum_{\alpha_i} \dot{S}^{\alpha_i}$ and therefore
\begin{align}
\dot{\mathcal{F}}_i=\dot{E}_i+T_i \dot{I}_i.
\end{align}
The local nonequilibrium free energy rate consists therefore of the energy-related and the information-related contribution. At the steady state the internal energy of the system is constant ($d_t E=0$) and thus $E_i$ can be interpreted as the energy flow to the subsystem $j \neq i$. The subsystem attached to an isothermal environment may therefore perform work either due to the energy flow from the other subsystem or due to the information flow; the latter case corresponds to the operation of information-powered devices.

\textit{Example}. The applicability of our approach will now be demonstrated on a recently proposed~\cite{ptaszynski2018} model of autonomous quantum Maxwell demon. Here we describe the device only briefly; for more details we refer to the original paper.

\begin{figure}
	\centering
	\includegraphics[width=0.95\linewidth]{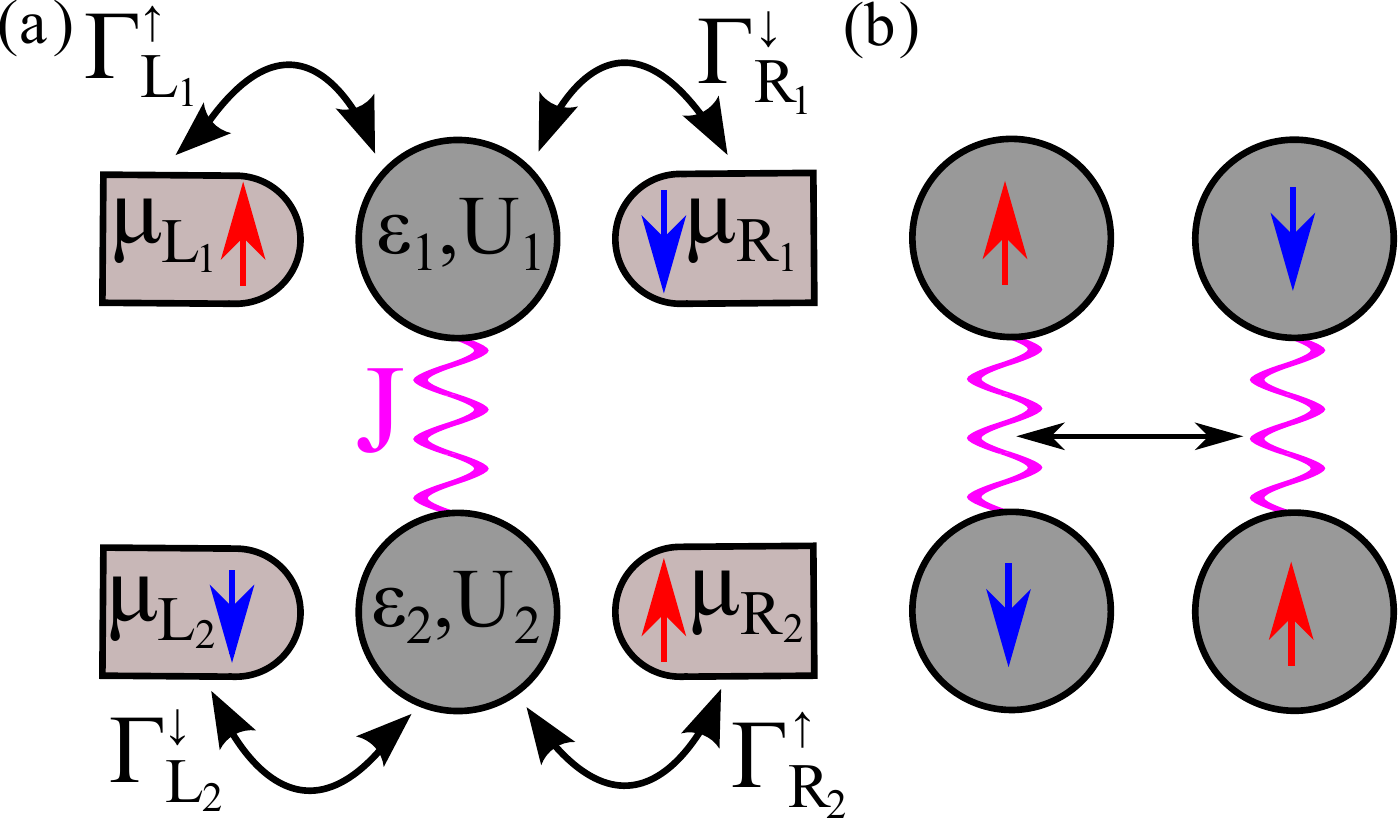}		
	\caption{(a) Scheme of the autonomous quantum Maxwell demon described in the main text. (b) Schematic representation of the spin exchange induced by the XY interaction.}
	\label{fig:demonschem}
\end{figure}

The analyzed setup [Fig.~\ref{fig:demonschem}~(a)] is composed of two quantum dots coupled by the XY exchange interaction, each attached to two electrodes with equal temperatures $T$. The Hamiltonian of the system is defined as
\begin{align}
\hat{H}_S =&\sum_{i \sigma} \epsilon_i d_{i\sigma}^\dagger d_{i\sigma} + \sum_{i} U_i n_{i \uparrow} n_{i \downarrow} \\ \nonumber &+\frac{J}{2} \left({d}_{1\uparrow}^\dagger {d}_{1\downarrow} d_{2\downarrow}^\dagger d_{2\uparrow} +{d}_{1\downarrow}^\dagger {d}_{1\uparrow} d_{2\uparrow}^\dagger d_{2\downarrow} \right),
\end{align}
where $d_{i\sigma}^\dagger$ ($d_{i\sigma}$) is the creation (annihilation) operator of an electron with spin $\sigma \in \{\uparrow, \downarrow\}$ in the dot $i \in \{1,2\}$, $n_{i \sigma}=d_{i\sigma}^\dagger d_{i\sigma}$ is the particle number operator, $\epsilon_i$ is the orbital energy, $U_i$ is the intra-dot Coulomb interaction in the dot $i$ and $J$ is the exchange coupling. The bath Hamiltonian reads $\hat{H}_B=\sum_{\alpha_i k \sigma} \epsilon_{\alpha_i k \sigma} c^\dagger_{\alpha_i k \sigma} c_{\alpha_i k \sigma}$, where $c^\dagger_{\alpha_i k \sigma}$ ($c^\dagger_{\alpha_i k \sigma}$) is the creation (annihilation) operator of an electron with spin $\sigma$, wave number $k$ and energy $\epsilon_{\alpha_i k \sigma}$ in the reservoir $\alpha_i$; here $\alpha_i=L_i$ ($R_i$) denotes the left (right) reservoir attached to the dot $i$. Finally, the system bath-interaction Hamiltonian is expressed as $\hat{H}_{I}=\sum_{i \alpha_i k \sigma} t_{\alpha_i} c^\dagger_{\alpha_i k \sigma} d_{i \sigma}+ \text{h.c.}$, where $t_{\alpha_i}$ is the tunnel coupling of the dot $i$ to the reservoir $\alpha_i$. We also define the coupling strength $\Gamma_{\alpha_i}^\sigma=2 \pi |t_{\alpha_i}|^2 \rho_{\alpha_i}^\sigma$, where $\rho_{\alpha_i}^\sigma$ is the density of states of electrons with spin $\sigma$ in the bath $\alpha_i$.
%
%

To describe the dynamics of the device we apply a microscopically derived Lindblad equation which couples populations to coherences and is thermodynamically consistent in the weak coupling limit (it is equivalent to the phenomenological approach proposed in Ref.~\cite{kirsanskas2018}). Details of the method, discussion of its limits of validity and comparison with the secular Lindblad equation (which is by construction thermodynamically consistent but neglects genuine quantum coherent effects) are presented in the Supplementary Material~\cite{supp}.

The device works as follows: The baths are taken to be fully spin polarized, i.e., either $\Gamma_{\alpha_i}^\uparrow$ or $\Gamma_{\alpha_i}^\downarrow$ is equal to 0. As a result, the electrodes act as spin filters which forbid the tunneling of electrons with a spin opposite to the polarization~\cite{rudzinski2001, braun2004}. Additionally, the polarizations of reservoirs attached to a single dot are arranged in an anti-parallel way, such that the electron may be transferred between the electrodes only when it changes its spin. This is enabled by the XY interaction which exchanges spins between the dots [Fig.~\ref{fig:demonschem}~(b)]. Since this interaction conserves the total spin, the spin flips occur simultaneously in both subsystems and thus the steady-state currents through both dots have to be equal. Let us now apply a high positive bias $V_1=\mu_{L_1}-\mu_{R_1}>0$ to the first dot, and a smaller opposite bias $V_2=\mu_{L_2}-\mu_{R_2}<0$ ($|V_2|<|V_1|$) to the other one. Then, the voltage-driven current through the first dot will pump electrons in the second dot against the bias, which is due to a nonequilibrium spin population induced by spin flips. This can be interpreted as the operation of a Maxwell demon: The high positive voltage in the first dot tends to reset the upper dot to the state $\uparrow$ (i.e., the singly occupied state with a spin up). As a result, the spin dynamics generated by the XY interaction (equivalent to the operation of the quantum iSWAP gate~\cite{schuch2003}) flips the spin in the second dot if it is in the state $\downarrow$, whereas leaves it unchanged when it is in the state $\uparrow$ [cf. Fig.~\ref{fig:demonschem}~(b)], thus creating an excess populations of spins $\uparrow$. This feedback mechanism induces the information flow between the dots, thus enabling a conversion of heat into work.

%
\begin{figure}
	\centering
	\includegraphics[width=0.94\linewidth]{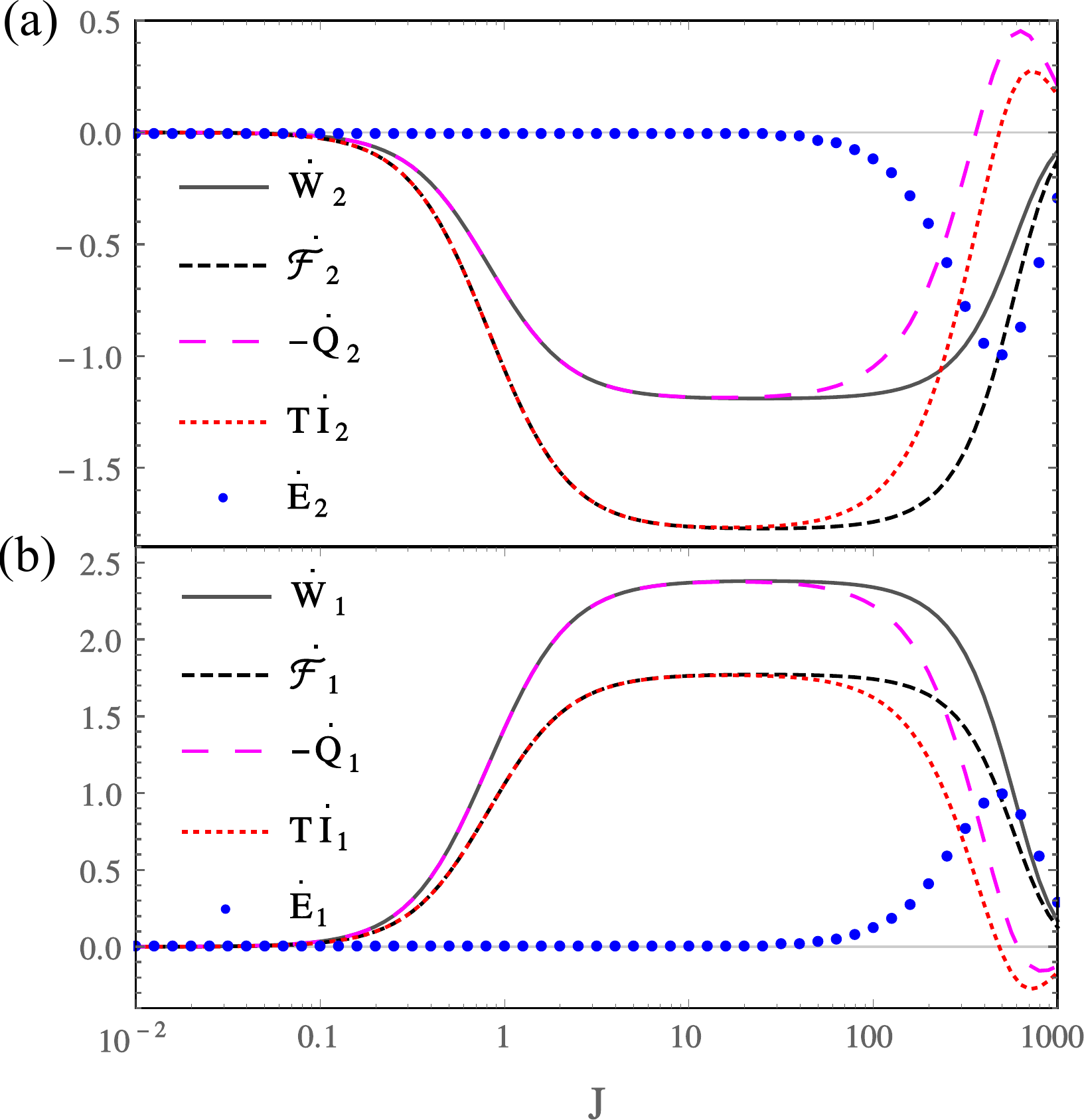}		
	\caption{Steady state work, free energy, heat, information flow and energy flow for the second (a) and the first (b) quantum dot. Results for $T=100$, $\epsilon_i=0$, $U_i \rightarrow \infty$ (strong Coulomb blockade), $\mu_{L_1}=-\mu_{R_1}=60$, $\mu_{L_2}=-\mu_{R_2}=-30$, $\Gamma_{L_1}^\downarrow=\Gamma_{R_1}^\uparrow=\Gamma_{L_1}^\uparrow=\Gamma_{R_2}^\downarrow=0$, all other coupling strengths $\Gamma_{\alpha_i}^\sigma$ equal to $\Gamma=1$.}
	\label{fig:demonres}
\end{figure}
%

Our local Clausius and free energy inequalities [Eqs.~\eqref{2ndlawloc},~\eqref{locfreeenineq}] for this system are demonstrated in Fig.~\ref{fig:demonres}. As one can see, for $J \lessapprox 500 \Gamma$ the dot $2$ cools its isothermal environment, which is enabled by the information flow (${T \dot{I}_2<-\dot{Q}_2<0}$); at the same time it performs work (i.e, pumps the current against the voltage), which is possible due to the negative free energy rate (${\dot{\mathcal{F}}_2<\dot{W}_2<0}$). This is compensated by the dissipation of work into heat in the first dot. We emphasize that for $J \lessapprox 50\Gamma$ the work is performed in the second dot only due to feedback-induced information flow ($\dot{\mathcal{F}}_2 \approx T \dot{I}_2 $) and not due to energy flow, which is negligible ($E_2 \approx 0$); this justifies the interpretation of the device as a Maxwell demon. As shown in the Supplementary Material~\cite{supp}, in this regime the information flow is generated by the unitary spin dynamics rather than the dissipative tunneling dynamics, in contrast to the classical Maxwell demon studied in Ref.~\cite{strasberg2013}. On the other hand, for $J \gtrapprox 50\Gamma$ one can observe a noticeable energy flow from the first to the second dot which results from the splitting of energy levels. As a consequence, for $J \gtrapprox 500\Gamma$ the dot 2 starts to heat its environment ($-\dot{Q}_2>0$) and the setup ceases to work as a Maxwell demon. However, the dot 2 still performs work thanks to the negativity of the nonequilibrium free energy rate $\dot{\mathcal{F}}_2$, which now includes a significant energy-related contribution $\dot{E}_2$.

\textit{Conclusions}. Our inequality~\eqref{freeenineq}, by providing a complement to the second law, may have novel implications for the design of quantum heat engines~\cite{benenti2017} which need to be explored. In turn, our inequalities~\eqref{2ndlawloc} and~\eqref{locfreeenineq} provide the basis for thermodynamics of quantum information processing. We hope that, as happened with the classical counterpart of Eq.~\eqref{2ndlawloc}, numerous applications and experiments~\cite{koski2015} will also ensure it in the quantum realm.

\begin{acknowledgments}
\textit{Acknowledgements}.
K. P. is supported by the National Science Centre, Poland, under the project Opus 11 (No.~2016/21/B/ST3/02160) and the doctoral scholarship Etiuda 6 (No.~2018/28/T/ST3/00154). M. E. is supported by the European Research Council project NanoThermo (ERC-2015-CoG
Agreement No. 681456).
\end{acknowledgments}

\end{document}


	
\title{Supplementary material to: Thermodynamics of Quantum Information Flows}
	
\author{Krzysztof Ptaszy\'{n}ski}
\affiliation{Institute of Molecular Physics, Polish Academy of Sciences, ul.~M.~Smoluchowskiego 17, 60-179 Pozna\'{n}, Poland}
\email{krzysztof.ptaszynski@ifmpan.poznan.pl}
\author{Massimiliano Esposito}
\affiliation{Physics and Materials Science Research Unit, University of Luxembourg, L-1511 Luxembourg, Luxembourg}
	
\date{\today}

\maketitle

\section*{Supplementary material}
This appendix contains in the following order:

\begin{itemize}
	\item General derivation of the non-secular Lindblad equation from the Redfield equation [Sec.~1]
	\item Derivation of the non-secular Lindblad equation for the autonomous Maxwell demon [Sec.~2]
	\item Discussion of the thermodynamic consistency of the non-secular Lindblad equation [Sec.~3]
	\item Comparison with the results obtained by means of the secular approximation [Sec.~4]
	\item Discussion of the non-bipartite structure of dynamics generated by the secular master equation [Sec.~5]
	\item Demonstration of the equivalence of Eq.~(11) from Ref.~\cite{horowitz2014} with Eq.~(18) from the main text for the case of bipartite dynamics [Sec.~6]
	\item Calculation of the unitary and the dissipative contribution to the information flow [Sec.~7]
\end{itemize}

\subsection{Derivation of the non-secular Lindblad equation}
In the main text it was assumed that one can describe the dynamics of the system by the Lindblad equation providing the Gibbs state at equilibrium. However, a most common approach to the dynamics of open quantum systems, namely, the perturbatively derived Redfield equation, is not of Lindblad form and thus does not guarantee the complete positivity of the dynamics~\cite{breuer2002}. Furthermore, it does not provide an exact Gibbs state, and thus may be inconsistent with the second law of thermodynamics~\cite{argentieri2014}. 

The non-completely positive dynamics arises due to the presence of terms in the Redfield equation coupling the state populations to the coherences between the non-degenerates eigenstates of the system (referred to as the non-secular terms). The usual way to ensure the complete positivity is to apply the secular approximation which neglects such terms. In such a way one obtains a thermodynamically consistent master equation of Lindblad form~\cite{breuer2002}. However, as a result one may miss some genuine quantum coherent effects, such as a finite time-scale of coherent oscillations between states of the system~\cite{palmieri2009, kirsanskas2018}.

To deal with such a problem, Refs.~\cite{palmieri2009} and~\cite{kirsanskas2018} proposed, on purely phenomenological grounds, a Lindblad equation including the non-secular terms. Here we provide a rigorous theoretical justification of such an approach by deriving a corresponding Lindblad equation directly from the Redfield equation. This is done by applying a set of approximations which modifies the form of non-secular terms instead of canceling them. We will first describe a general framework; the derivation of the master equation for the autonomous Maxwell demon analyzed in the main text is presented in the next section. 

Following Ref.~\cite{breuer2002}, we express a generic form of the interaction Hamiltonian as
\begin{align} \label{hamint}
\hat{H}_I = \sum_{\alpha} A_\alpha \otimes B_\alpha,
\end{align}
where $A_\alpha$ and $B_\alpha$ are the operators acting in the Hilbert space of the system and the reservoirs, respectively. After applying the Born-Markov approximation, dynamics of the density matrix of the system in the interaction picture is given by the Redfield equation~\cite{breuer2002}
\begin{align}
&d_t \rho_I (t)=\sum_{\omega, \omega'} \sum_{\alpha, \beta} e^{i (\omega'-\omega)} C_{\alpha \beta} (\omega) \\ \nonumber & \times \left[ A_\beta (\omega) \rho_I (t) A_{\alpha}^\dagger (\omega') - A_{\alpha}^\dagger (\omega') A_\beta (\omega) \rho_I (t) \right]+\text{h.c.}
\end{align}
Here
\begin{align}
A_\alpha(\omega)= \sum_{E_l-E_k=\omega} |l \rangle \langle l| A_\alpha |k \rangle \langle k|,
\end{align}
are the operators describing jumps between eigenstates $|k\rangle$ and $|l \rangle$ of the Hamiltonian $\hat{H}_S$ differing by the energy gap ${\omega=E_l-E_k}$. The coefficients
\begin{align} \label{fourier}
C_{\alpha \beta} (\omega)= \int_{0}^{\infty} d\tau e^{i \omega \tau} \text{Tr}_B \left[B_\alpha^\dagger(\tau) B_\beta(0) \rho_B \right],
\end{align}
are the one sided Fourier transforms of the bath correlation functions. Here $\text{Tr}_B$ denotes the partial trace over degrees of freedom of the reservoirs, $\rho_B$ is the density matrix of the bath and 
\begin{align}
B_\alpha (\tau)=e^{i \hat{H}_B \tau} B_\alpha e^{-i \hat{H}_B \tau},
\end{align}
are the interaction picture operators of the environment. 

The Redfield equation can be rewritten in the Schr\"{o}dinger picture as
\begin{align} \label{redschrod}
& d_t \rho (t)=-i \left[\hat{H}_S, \rho(t) \right]  \\ \nonumber&+\sum_{\omega, \omega'} \sum_{\alpha, \beta} C_{\alpha \beta} (\omega) \left[ A_\beta (\omega) \rho (t) A_{\alpha}^\dagger (\omega') - A_{\alpha}^\dagger (\omega') A_\beta (\omega) \rho (t) \right] \\ \nonumber &+\sum_{\omega, \omega'} \sum_{\alpha, \beta} C_{\beta \alpha}^* (\omega) \left[ A_\alpha (\omega') \rho (t) A_{\beta}^\dagger (\omega) - \rho(t) A_{\beta}^\dagger (\omega) A_\alpha (\omega') \right].
\end{align}
One can also express the Fourier transform of the bath correlation functions as a sum of their real and imaginary part
\begin{align}
C_{\alpha \beta} (\omega) = \frac{1}{2} \gamma_{\alpha \beta} (\omega)+i S_{\alpha \beta} (\omega),
\end{align}
with coefficients $\gamma_{\alpha \beta} (\omega)$ being positive and $S_{\alpha \beta} (\omega)$ forming a real symmetric matrix~\cite{breuer2002}.

Within the secular approximation one assumes that a typical time of intrinsic evolution of the system is much smaller then the relaxation time of the system $\tau_R$: $|\omega-\omega'|^{-1} \ll \tau_R$. One can therefore neglect the rapidly oscillating terms of the Redfield equation by taking $A_{\alpha} (\omega' \neq \omega)=0$. Here, following similar way of thinking, we assume that for $|\omega-\omega'|^{-1} \ll \tau_R$ the non-secular terms of the Redfield equation corresponding to $\omega=\omega'$ does not influence much the dynamics of the populations, such that they may be kept into the master equation; moreover, their exact form is not very important and thus may be modified without much consequences. We also assume that the functions $C_{\alpha \beta} (\omega)$ are smooth, i.e., $C_{\alpha \beta} (\omega) \approx C_{\alpha \beta} (\omega')$ for $|\omega-\omega'|^{-1} \approx \tau_R$. One may then replace
\begin{align}
\gamma_{\alpha \beta}(\omega), \gamma_{\alpha \beta}(\omega') \rightarrow \gamma_{\alpha \beta} (\omega,\omega'), \\
S_{\alpha \beta}(\omega), S_{\alpha \beta}(\omega') \rightarrow S_{\alpha \beta}(\omega, \omega'),
\end{align}
where
\begin{align}
\gamma_{\alpha \beta}(\omega, \omega') = \sqrt{\gamma_{\alpha \beta}(\omega) \gamma_{\alpha \beta}(\omega')  }, \\
S_{\alpha \beta}(\omega, \omega') = \frac{ S_{\alpha \beta} (\omega) +S_{\alpha \beta} (\omega')}{2},
\end{align}
such that  $\gamma_{\alpha \beta}(\omega) \approx \gamma_{\alpha \beta}(\omega') \approx \gamma_{\alpha \beta} (\omega,\omega')$ and $S_{\alpha \beta}(\omega) \approx S_{\alpha \beta}(\omega') \approx S_{\alpha \beta}(\omega, \omega')$ for $|\omega-\omega'|^{-1} \approx \tau_R$. One can now perform the following replacement in Eq.~\eqref{redschrod}:
\begin{align}
C_{\alpha \beta} (\omega) \rightarrow \frac{1}{2} \gamma_{\alpha \beta} (\omega, \omega')+i S_{\alpha \beta} (\omega,\omega'), \\
C_{\beta \alpha}^* (\omega') \rightarrow \frac{1}{2} \gamma_{\beta \alpha} (\omega, \omega')-i S_{\beta \alpha} (\omega,\omega').
\end{align}
Changing indexes $\alpha \leftrightarrow \beta$ and $\omega \leftrightarrow \omega'$ in the second sum in Eq.~\eqref{redschrod} one obtains
\begin{align} \label{lindgen} \nonumber 
&d_t \rho (t)=-i \left[\hat{H}_S+\hat{H}_{LS}, \rho(t) \right]+\sum_{\omega,\omega'} \sum_{\alpha,\beta} \gamma_{\alpha \beta} (\omega,\omega') \\ &\times \left[ A_\beta (\omega) \rho (t) A_{\alpha}^\dagger (\omega') - \frac{1}{2}\left\{ A_{\alpha}^\dagger (\omega') A_\beta (\omega), \rho (t) \right\} \right],
\end{align}
where the last part of the equation defines the dissipator $\mathcal{D}$. Here
\begin{align}
\hat{H}_{LS}=\sum_{\omega,\omega'} \sum_{\alpha,\beta} S_{\alpha \beta}(\omega, \omega') A_{\alpha}^\dagger (\omega') A_\beta (\omega),
\end{align}
is the Lamb shift Hamiltonian commuting with $\hat{H}_S$. The commutation property can be verified by taking into account the explicit form of operators $A_\alpha$ and expressing $\hat{H}_{LS}$ as
\begin{align}
\hat{H}_{LS} = &\sum_{klmn} \sum_{\alpha,\beta} S_{\alpha \beta}(\omega_{mn}, \omega_{lk}) \\ \nonumber &\times |n \rangle \langle n| A_{\alpha}^\dagger |m \rangle \langle m|l \rangle \langle l| A_\beta|k \rangle \langle k| \\ \nonumber
= &\sum_{kln} \sum_{\alpha,\beta} S_{\alpha \beta}(\omega_{ln}, \omega_{lk}) |n \rangle \langle n| A_{\alpha}^\dagger |l \rangle \langle l| A_\beta|k \rangle \langle k|.
\end{align}
The commutator can be then expressed as
\begin{align}
&\left[\hat{H}_S, \hat{H}_{LS} \right] \\ \nonumber
&=\sum_{kln} \sum_{\alpha,\beta} S_{\alpha \beta}(\omega_{ln}, \omega_{lk}) E_n |n \rangle \langle n| A_{\alpha}^\dagger |l \rangle \langle l| A_\beta|k \rangle \langle k| \\ \nonumber
&-\sum_{kln} \sum_{\alpha,\beta} S_{\alpha \beta}(\omega_{ln}, \omega_{lk}) E_k |n \rangle \langle n| A_{\alpha}^\dagger |l \rangle \langle l| A_\beta|k \rangle \langle k|.
\end{align}
Since each pair $E_k, E_l$ will appear two times and ${E_l-E_k}={-(E_k-E_l)}$ the terms will cancel each other, and thus $[\hat{H}_S, \hat{H}_{LS}]=0$.

Equation~\eqref{lindgen} is of the first standard form, and thus ensures the completely-positive trace-preserving dynamics~\cite{breuer2002}. Since the matrix $\gamma_{\alpha \beta} (\omega,\omega')$ is positive, the equation can be brought to the Lindblad form by diagonalizing the matrix $\gamma_{\alpha \beta} (\omega,\omega')$. We will demonstrate this in the next subsection for a particular case.

\subsection{Lindblad equation for the Maxwell demon}
Following the general discussion from the previous subsection, let us now derive the Lindblad equation for the autonomous Maxwell demon considered in the main text. Let us first note, that in Eq.~\eqref{hamint} it was assumed that the interaction Hamiltonian can be expressed as a tensor product of operators acting on the Hilbert spaces of the system and the bath; this requires that the operators $A_\alpha$ and $B_\alpha$ commute. On the other hand, in the considered system creation and annihilation operators included in the interaction Hamiltonian 
\begin{align}
\hat{H}_{I}=\sum_{i \alpha_i k \sigma} t_{\alpha_i} c^\dagger_{\alpha_i k \sigma} d_{i \sigma}+ t_{\alpha_i}^* c_{\alpha_i k \sigma} d^\dagger_{i \sigma},
\end{align}
follow the fermionic anticommutation relations. However, this problem can be solved by employing the Jordan-Wigner transformation~\cite{schaller2014, schaller2009}. Let us first replace each set of quantum numbers $i\sigma$ or $\alpha_i k \sigma$ with a single quantum number: $i\sigma \rightarrow \kappa$, $\alpha_i k \sigma \rightarrow \lambda$, with $\kappa \in \{1,K\}$ and $\lambda \in \{1,L\}$. Annihilation operators can be then represented by tensor products of Pauli matrices
\begin{align}
d_\kappa &=\prod_{\kappa'=1}^{\kappa-1} \sigma^z_{\kappa'} \otimes \sigma_\kappa^- \otimes \prod_{{\kappa''}=\kappa+1}^{K} I_{\kappa''} \otimes \prod_{\lambda=1}^L I_{\lambda}, \\
c_\lambda &= \prod_{\kappa=1}^K \sigma^z_{\kappa} \otimes \prod_{\lambda'=1}^{\lambda-1} \sigma^z_{\lambda'} \otimes \sigma_\lambda^- \otimes \prod_{{\lambda''}=\lambda+1}^{L} I_{\lambda''},
\end{align}
where $\sigma^-=(\sigma^z-i \sigma^y)/2$ is the lowering operator and products $\Pi$ are defined as tensor products
\begin{align}
\prod_{n=a}^{b} \sigma_n^j=\sigma_a^j \otimes \sigma_{a+1}^j \otimes ... \otimes \sigma_{b}^j.
\end{align}
Creation operators are defined by replacing the lowering operator $\sigma^-$ with the raising operator $\sigma^+=(\sigma^-)^\dagger=(\sigma^z+i \sigma^y)/2$. One can verify, that the operators defined in such a way follow the fermionic anicommutation relations.

Let us now introduce the modified operators
\begin{align}
\tilde{d}_\kappa &=-\prod_{\kappa'=1}^{\kappa-1} I_{\kappa'} \otimes \sigma_\kappa^- \otimes \prod_{{\kappa''}=\kappa+1}^{K} \sigma^z_{\kappa''} \otimes \prod_{\lambda=1}^L I_{\lambda}, \\
\tilde{c}_\lambda &= \prod_{\kappa=1}^K I_{\kappa} \otimes \prod_{\lambda'=1}^{\lambda-1} \sigma^z_{\lambda'} \otimes \sigma_\lambda^- \otimes \prod_{{\lambda''}=\lambda+1}^{L} I_{\lambda''},
\end{align}
which are obtained from $d_\kappa$ and $c_\lambda$ by replacing $\sigma^z_{\kappa'} \rightarrow I_{\kappa'}$, $I_{\kappa''} \rightarrow {\sigma^z_{\kappa''}}$ and $\sigma^-_\kappa \rightarrow - \sigma^-_\kappa$. These operators may be interpreted as creation and annihilation operators which act separately on the Hilbert spaces of the system and the bath, and therefore follow the commutation relations such as $[\tilde{d}_\kappa,\tilde{c}_\lambda ]=0$. Their fermionic character is retained due to the fact that pairs of operators acting on either the system or the bath still anticommute; for example $\{\tilde{d}_\kappa,\tilde{d}_{\kappa'} \}=0$. One may then verify, that the Hamiltonian will not change upon replacing $d_\kappa, c_\lambda \rightarrow \tilde{d}_\kappa, \tilde{c}_\lambda$. Let us demonstrate this on the example of a pair of operators $c_\lambda^\dagger d_\kappa$ appearing in the interaction Hamiltonian
\begin{align}
c_\lambda^\dagger d_\kappa &= \prod_{\kappa'=1}^{\kappa-1} (\sigma^z_{\kappa'})^2 \otimes \sigma_\kappa^z \sigma_\kappa^+ \otimes \prod_{{\kappa''}=\kappa+1}^{K} \sigma^z_{\kappa''} I_{\kappa''} \\ \nonumber
&\otimes \prod_{\lambda'=1}^{\lambda-1} \sigma^z_{\lambda'} I_{\lambda'} \otimes \sigma_\lambda^- I_\lambda \otimes \prod_{{\lambda''}=\lambda+1}^{L} I_{\lambda''} I_{\lambda''} \\ \nonumber
&=\prod_{\kappa'=1}^{\kappa-1} (I_{\kappa'})^2 \otimes (-I_\kappa \sigma_\kappa^+) \otimes \prod_{{\kappa''}=\kappa+1}^{K} I_{\kappa''} \sigma^z_{\kappa''} \\ \nonumber
&\otimes \prod_{\lambda'=1}^{\lambda-1} \sigma^z_{\lambda'} I_{\lambda'} \otimes \sigma_\lambda^- I_\lambda \otimes \prod_{{\lambda''}=\lambda+1}^{L} I_{\lambda''} I_{\lambda''} = \tilde{c}_\lambda^\dagger \tilde{d}_\kappa,
\end{align}
where we have used the properties $(\sigma^z)^2=I$ and $\sigma_z \sigma^\pm=-\sigma^\pm$.

The interaction Hamiltonian can be therefore rewritten as
\begin{align}
\hat{H}_I = \sum_{\alpha_i \sigma} \sum_{+,-} A_{\alpha_i \sigma \pm} \otimes B_{i\alpha_i  \sigma \pm},
\end{align}
where
\begin{align}
A_{\alpha_i \sigma +} &=\tilde{d}_{i \sigma}^\dagger, \\
A_{\alpha_i \sigma -} &=\tilde{d}_{i \sigma}, \\
B_{\alpha_i \sigma +} &=\sum_{k} \tilde{c}_{\alpha_i k \sigma} t_{\alpha_i}^*, \\
B_{\alpha_i \sigma -} &=\sum_{k} \tilde{c}_{\alpha_i k \sigma}^\dagger t_{\alpha_i}.
\end{align}

In the considered system, the reduced density matrix of the bath can be written as
\begin{align}
\rho_B = \prod_{\alpha_i} Z_{\alpha_i}^{-1} e^{-\beta_{\alpha_i} (\hat{H}_{\alpha_i} - \mu_{\alpha_i} \hat{N}_{\alpha_i})},
\end{align}
where
\begin{align}
\hat{H}_{\alpha_i} &= \sum_{k \sigma} \epsilon_{\alpha_i k \sigma} c^\dagger_{\alpha_i k \sigma} c_{\alpha_i k \sigma}, \\
\hat{N}_{\alpha_i} &= \sum_{k \sigma} c^\dagger_{\alpha_i k \sigma} c_{\alpha_i k \sigma},
\end{align}
are the Hamiltonian and the particle number operator of the reservoir $\alpha_i$, respectively, and $Z_{\alpha_i}={\text{Tr} \{\exp[-\beta_{\alpha_i} (\hat{H}_{\alpha_i} - \mu_{\alpha_i} N_{\alpha_i})]\}}$ is the partition function. One may then verify that the Fourier transform of the bath correlation function defined by Eq.~\eqref{fourier} vanishes for $\alpha \neq \beta$. In the continuous limit, one may also replace trace in the definition of the bath correlation functions by an integral. As a result, the Redfield equation in the Schr\"{o}dinger picture [Eq.~\eqref{redschrod}] can be expressed as
\begin{align}
& d_t \rho (t)=-i \left[\hat{H}_S, \rho \right]  \\ \nonumber &+ \left\{ \sum_{\omega, \omega'} \sum_{i \alpha_i \sigma} \sum_{+,-} C_{\alpha_i \sigma \pm} (\omega) \times \left[ A_{\alpha_i \sigma \pm} (\omega) \rho A_{\alpha_i \sigma \pm}^\dagger (\omega') \right. \right. \\ \nonumber 
& \left. \left. - A_{\alpha_i \sigma \pm}^\dagger (\omega') A_{\alpha_i \sigma \pm} (\omega) \rho \right] + \text{h.c.} \right\},
\end{align}
where 
\begin{align}
A_{\alpha_i \sigma \pm}(\omega)= \sum_{E_l-E_k=\omega} |l \rangle \langle l| A_{\alpha_i \sigma \pm} |k \rangle \langle k|,
\end{align}
are the jump operators, and the Fourier transforms of the bath correlations functions read as
\begin{align}
&C_{\alpha_i \sigma \pm}(\omega)= \\ \nonumber &\int \displaylimits_{0}^{\infty} d\tau e^{i\omega \tau} \int \displaylimits_{-\infty}^\infty dE \rho_{\alpha_i}^\sigma (E) |t_{\alpha_i}|^2 e^{\mp iE\tau} f\left[\pm \beta_{\alpha_i} \left(E-\mu_{\alpha_i} \right) \right],
\end{align}
where $\rho_{\alpha_i}^\sigma (E)$ is the density of states in the lead $\alpha_i$ for the spin $\sigma$. We will further assume that $\rho_{\alpha_i}^\sigma (E)=\text{const.}$ and define the coupling strength $\Gamma_{\alpha_i}^\sigma=2 \pi |t_{\alpha_i}|^2 \rho_{\alpha_i}^\sigma$. The function $C_{\alpha_i \sigma \pm}(\omega)$ reads then
\begin{align}
C_{\alpha_i \sigma \pm}(\omega)=\frac{1}{2} \gamma_{\alpha_i \sigma \pm}(\omega)+iS_{\alpha_i \sigma \pm}(\omega),
\end{align}
where
\begin{align}
\gamma_{\alpha_i \sigma \pm}(\omega) = \Gamma_{\alpha_i}^\sigma f\left[\beta_{\alpha_i} (\omega \mp \mu_{\alpha_i}) \right],
\end{align}
is the tunneling rate and
\begin{align}
S_{\alpha_i \sigma \pm}(\omega) &= \frac{\Gamma_{\alpha_i}^\sigma}{2 \pi} \mathcal{P} \int \displaylimits_{-\infty}^\infty \frac{dE f\left[\beta_{\alpha_i} (E \mp \mu_{\alpha_i}) \right]}{\omega-E} \\ \nonumber
&=-\frac{\Gamma_{\alpha_i}^\sigma}{2 \pi} \text{Re} \Psi \left[ \frac{1}{2} + i \frac{\beta_{\alpha_i} (\omega \mp \mu_{\alpha_i})}{2 \pi} \right],
\end{align}
is the principal part of the Cauchy integral of the tunneling rate describing the level renormalization~\cite{schaller2014, wunsch2005, splettstoesser2012}; here $\Psi$ denotes the digamma function.

Following the approach presented in the previous section, one obtains the non-secular Lindblad equation
\begin{align} \label{linddem}
d_t \rho=-i \left[\hat{H}_S+\hat{H}_{LS}, \rho \right]+ \sum_{i} \mathcal{D}_i \rho,
\end{align}
where the dissipator $\mathcal{D}_i$ is defined as
\begin{align} \label{dissloc} 
&\mathcal{D}_i \rho=  \sum_{\omega,\omega'} \sum_{\alpha_i \sigma} \sum_{+,-} \gamma_{\alpha_i \sigma \pm} (\omega,\omega') \\ \nonumber &\times \left[ A_{\alpha_i \sigma \pm} (\omega) \rho A_{\alpha_i \sigma \pm}^\dagger (\omega') - \frac{1}{2}\left\{ A_{\alpha_i \sigma \pm}^\dagger (\omega') A_{\alpha_i \sigma \pm} (\omega), \rho \right\} \right],
\end{align}
with $\gamma_{\alpha_i \sigma \pm} (\omega,\omega')= \sqrt{\gamma_{\alpha_i \sigma \pm} (\omega)\gamma_{\alpha_i \sigma \pm} (\omega')}$, and the Lamb shift Hamiltonian reads
\begin{align} \label{lamb} \nonumber
\hat{H}_{LS} =&\frac{1}{2 }\sum_{\omega,\omega'}  \sum_{i \alpha_i \sigma} \sum_{+,-} \left[S_{\alpha_i \sigma \pm}(\omega)+S_{\alpha_i \sigma \pm}(\omega') \right] \\ & \times A_{\alpha_i \sigma \pm}^\dagger (\omega') A_{\alpha_i \sigma \pm} (\omega).
\end{align}
The secular master equation is obtained if one further takes ${\gamma_{\alpha_i \sigma \pm} (\omega,\omega')=0}$ in Eq.~\eqref{dissloc}  and ${A_{\alpha_i \sigma \pm}^\dagger (\omega') A_{\alpha_i \sigma \pm} (\omega)=0}$ in Eq.~\eqref{lamb} for $\omega' \neq \omega$.

The dissipator $\mathcal{D}_i$ can be rewritten directly in the Lindblad form
\begin{align} \label{disfin}
\mathcal{D}_i \rho=\sum_{\alpha_i \sigma} \sum_{+,-} \left( L_{\alpha_i \sigma \pm} \rho L_{\alpha_i \sigma \pm}^\dagger - \frac{1}{2}\left\{ L_{\alpha_i \sigma \pm}^\dagger L_{\alpha_i \sigma \pm}, \rho \right\} \right),
\end{align}
with the Lindblad operators defined as
\begin{align}
L_{\alpha_i \sigma \pm}=\sum_{\omega} \sqrt{\gamma_{\alpha_i \sigma \pm} (\omega)} A_{i \alpha_i \sigma \pm}(\omega).
\end{align}
Explicitly, the Lindblad operators read as
\begin{align}
L_{\alpha_i \sigma+} &= \sum_{kl} \sqrt{\Gamma_{\alpha_i}^\sigma f^+_{\alpha_i} (\omega_{kl})} |k \rangle \langle k| c^\dagger_{i \sigma} | l \rangle \langle l|, \\
L_{\alpha_i \sigma-} &= \sum_{kl} \sqrt{\Gamma_{\alpha_i}^\sigma f^-_{\alpha_i} (\omega_{kl})} |k \rangle \langle k| c_{i \sigma} | l \rangle \langle l|,
\end{align}
where $f^\pm_{\alpha_i}(\omega)=f [\beta_{\alpha_i} (\omega \mp \mu_{\alpha_i})]$ and $\omega_{kl}=E_k-E_l$. As this form clearly illustrates, the operators describe the jumps between different superposition of eigenstates of the system with probability amplitudes of related to the tunneling rates. We note, that the same form of the Lindblad operators may be obtained by using the phenomenological approach proposed in Ref.~\cite{kirsanskas2018}.

\subsection{Relaxation to equilibrium}
%
\begin{figure}
	\centering
	\includegraphics[width=0.98\linewidth]{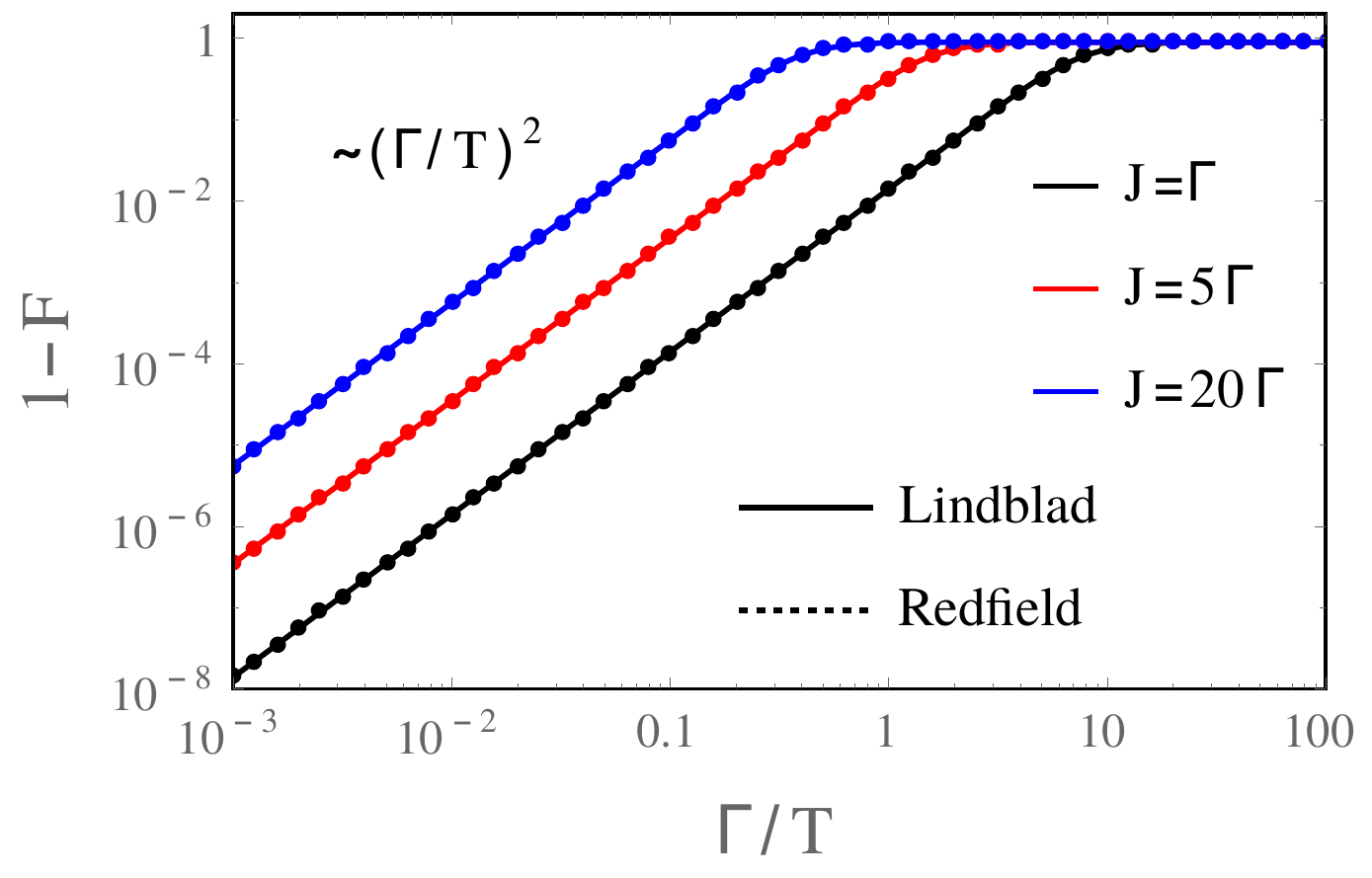}		
	\caption{Deviation from the Gibbs state $1-F$ for the non-secular Lindblad equation (solid lines) and the Redfield equation (dotted lines) for $\Gamma_{L_1}^\downarrow=\Gamma_{R_1}^\uparrow=\Gamma_{L_1}^\uparrow=\Gamma_{R_2}^\downarrow=0$, all other coupling strengths $\Gamma_{\alpha_i}^\sigma$ equal to $\Gamma$, $\epsilon_i=0$, $\mu_{\alpha_i}=0$ and $U_i \rightarrow \infty$ (strong Coulomb blockade).}
	\label{fig:fidel}
\end{figure}
%
The thermodynamic formalism described in the main text was based on the assumption, that the Gibbs state is a stationary state at equilibrium, i.e., for $\beta_\alpha=\beta$, $\mu_\alpha=\mu$. This is ensured by the Redfield equation within the secular approximation~\cite{breuer2002}. Unfortunately, this is not by construction guaranteed by the newly derived non-secular Lindblad equation (the similar observation has been already made for a phenomenological approach from Ref.~\cite{kirsanskas2018}). However, the steady state converges to the Gibbs state when the coupling to the bath is weak. We verify this numerically on the example of the autonomous quantum Maxwel demon. In particular, we calculate the convergence to the Gibbs state characterized by the fidelity defined as~\cite{nielsen2010}
\begin{align}
F=\left[ \text{Tr} \sqrt{\sqrt{\rho_{\text{eq}}} \rho_\text{st} \sqrt{\rho_{\text{eq}}}} \right]^2,
\end{align}
where $\rho_\text{st}$ denotes the stationary state of the master equation and the density matrix of the Gibbs state is defined as
\begin{align}
\rho_\text{eq} = Z_{\beta,\mu}^{-1} e^{-\beta \left(\hat{H}_S-\mu \hat{N} \right)},
\end{align}
with $Z_{\beta,\mu}=\text{Tr} \{\exp[ -\beta (\hat{H}_S-\mu \hat{N})]\}$ being the partition function. The fidelity takes values within the interval $[0,1]$ with $F=1$ indicating the perfect convergence of the density matrices~\cite{nielsen2010}. As shown in Fig.~\ref{fig:fidel} the fidelity converges to 1 in the weak coupling regime with the divergence from the Gibbs state scaling as $1-F \propto (\Gamma/T)^2$ for $\Gamma \lessapprox T$. As one can also observe, deviation from the Gibbs state is not a result of approximations made when deriving Lindblad equation from the Redfield equation, but is rather inherent to the perturbative nature of the Redfield equation itself.

\subsection{Comparison with the secular approximation}
\begin{figure}
	\centering
	\includegraphics[width=0.9\linewidth]{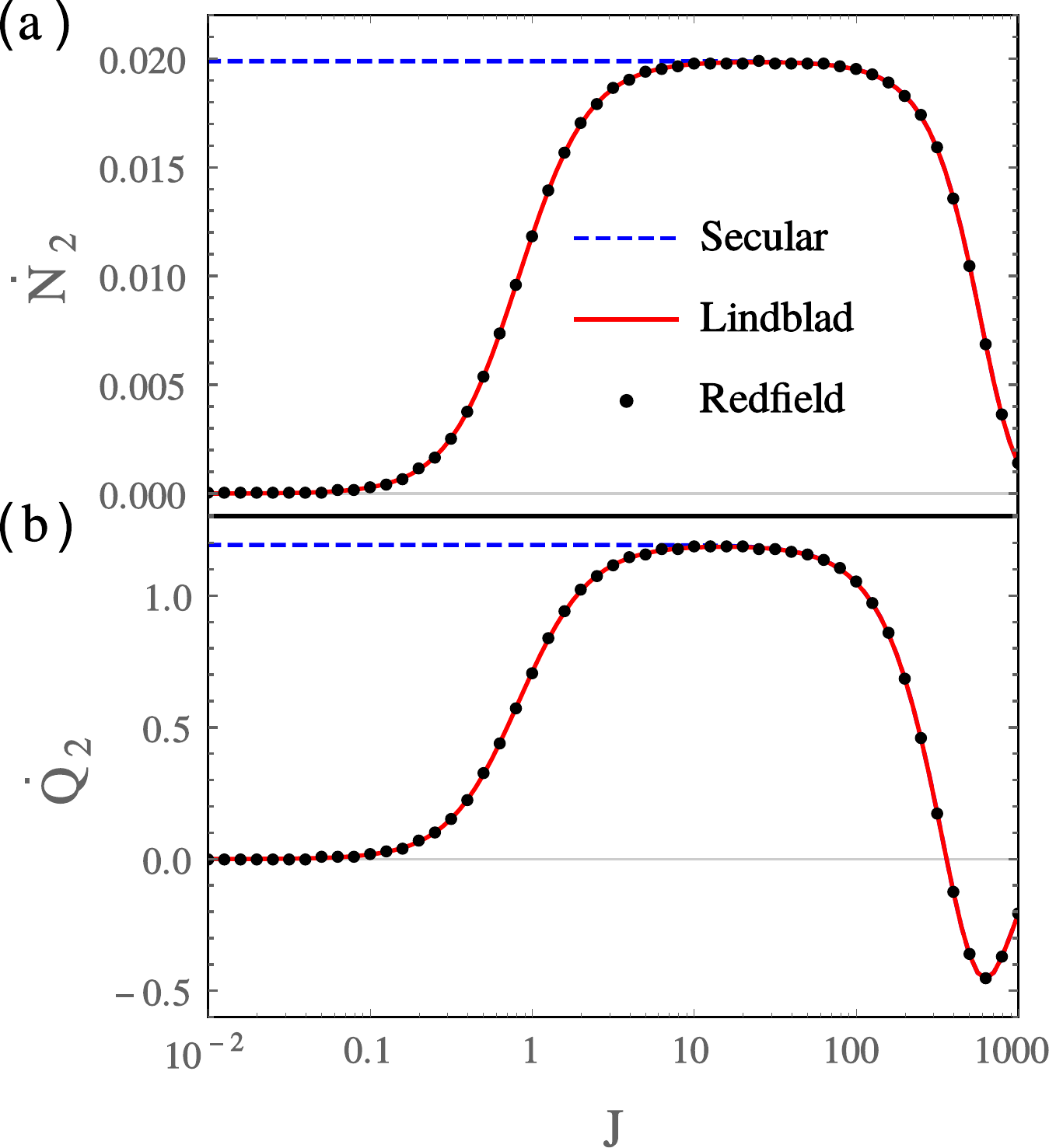}		
	\caption{Particle current (a) and cooling power (b) in the second dot calculated with the secular approximation (blue dashed line), non-secular Lindblad equation (red solid line) and the Redfield equation (black dots) for $T=100$, $\epsilon_i=0$, $U_i \rightarrow \infty$ (strong Coulomb blockade), $\mu_{L_1}=-\mu_{R_1}=60$, $\mu_{L_2}=-\mu_{R_2}=-30$, $\Gamma_{L_1}^\downarrow=\Gamma_{R_1}^\uparrow=\Gamma_{L_1}^\uparrow=\Gamma_{R_2}^\downarrow=0$, all other coupling strengths $\Gamma_{\alpha_i}^\sigma$ equal to $\Gamma=1$.}
	\label{fig:compsec}
\end{figure}

\begin{figure}
	\centering
	\includegraphics[width=0.94\linewidth]{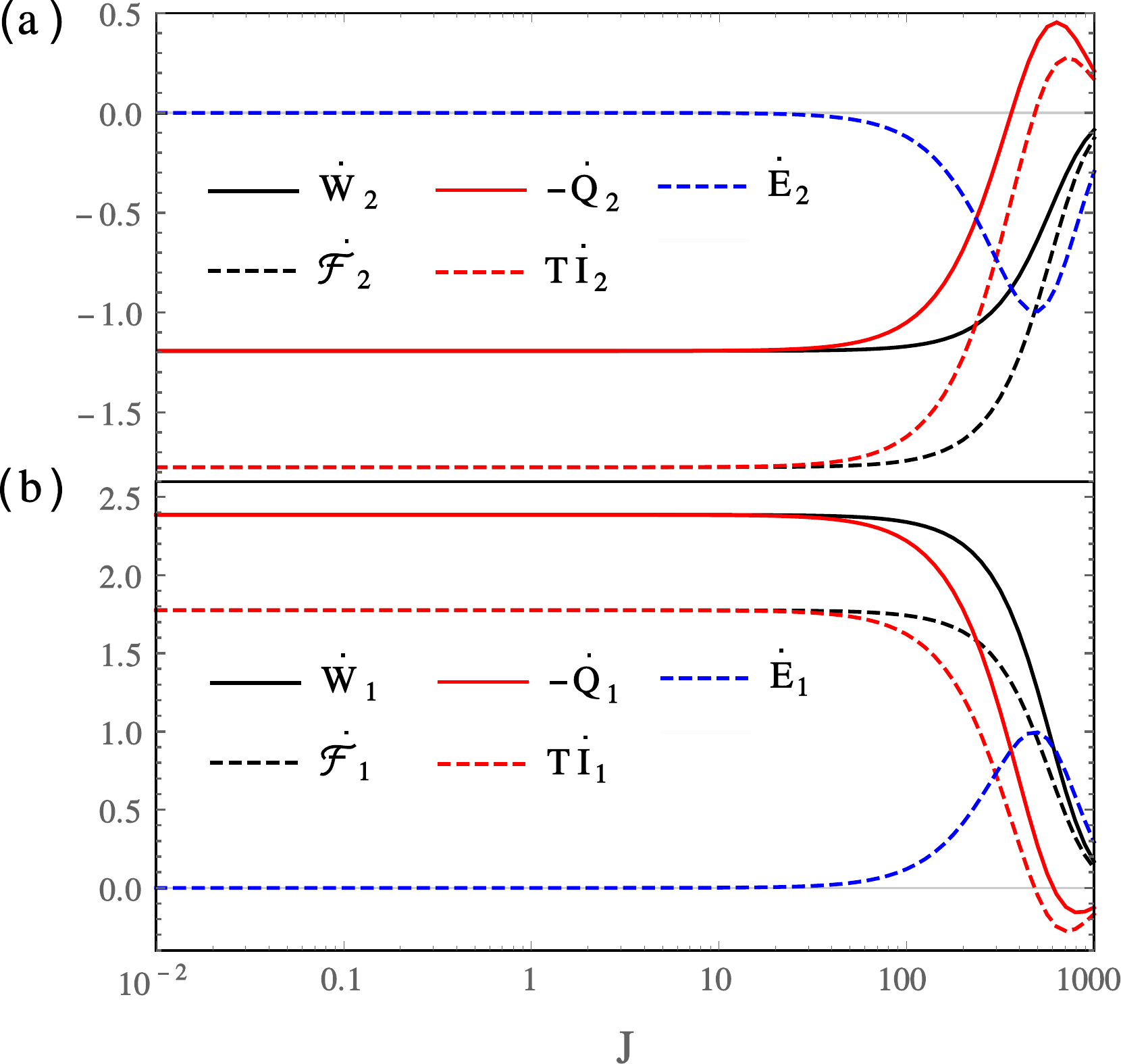}		
	\caption{Steady state work, free energy, heat, information flow and energy flow for the second (a) and the first (b) quantum dot calculated within the secular approximation. Parameters as in Fig.~\ref{fig:compsec}.}
	\label{fig:secinf}
\end{figure}

The thermodynamic consistency is by construction provided by the Redfield equation within the secular approximation. However, except from the case when the eigenstates are fully degenerate (analyzed in Ref.~\cite{cuetara2016}), the secular approximation decouples the populations from the coherences and thus neglects the genuine quantum coherent effects. This is demonstrated in Fig.~\ref{fig:compsec}. As shown, the secular master equation agrees with both the Redfield equation and the non-secular Lindblad equation for $J \gg \Gamma$. In this regime the assumption that the characteristic time of the intrinsic evolution of the system is much smaller than the relaxation time is valid. On the other hand, the secular approximation diverges from the other approaches when the ratio $J/\Gamma$ is relatively small. In particular, it predicts non-vanishing particle and heat currents for $J \rightarrow 0$, when the quantum dots become decoupled, which is clearly a non-physical result. This is a consequence of the fact that the secular approximation neglects the finite timescale of the coherent oscillations between the spin states. With this qualification, as demonstrated in Fig.~\ref{fig:secinf} the secular master equation can also describe the operation of the system as an autonomous Maxwell demon (cf.~Fig.~2 in the main text).

%
%
%
%

\subsection{Non-bipartite structure of the secular master equation}
We emphasize, that in the system we consider the secular approximation generates a classical rate equation describing the dynamics of eigenstate population, which, however, does not possess a bipartite structure as defined in Ref.~\cite{horowitz2014}. The term ``bipartite dynamics'' refer here to the situation when there are no transition generating the simultaneous change of states of both subsystems. As a result, the information flow cannot be analyzed within the approach proposed in Ref.~\cite{horowitz2014} and can only be described by our method. This clearly demonstrates the wider range of applicability of our approach even when the dynamics is effectively classical.
	
In the system we consider, the non-bipartite structure of the dynamics results from the fact that the total Hamiltonian $\hat{H}_S$ [Eq.~(21) in the main text] does not commute with the Hamiltonian of a single dot $\hat{H}_i=\epsilon_i d_{i\sigma}^\dagger d_{i\sigma} + U_i n_{i \uparrow} n_{i \downarrow}$ and thus the eigenstates of the total system cannot be expressed as products of states of the subsystems. More specifically, there exist eigenstates of the form $|+ \rangle ={( \alpha c_{1\uparrow}^\dagger c_{2 \downarrow}^\dagger + \beta c_{1\downarrow}^\dagger c_{2 \uparrow}^\dagger ) |0 \rangle} $ and $ |- \rangle ={( \beta c_{1\uparrow}^\dagger c_{2 \downarrow}^\dagger - \alpha c_{1\downarrow}^\dagger c_{2 \uparrow}^\dagger ) |0 \rangle}$ (with $|0 \rangle$ being the empty state) which are the entangled states of the first and the second dot. This leads to the presence of non-local correlations between the subsystems; for example, tunneling from one dot acts as a quantum measurement which ``fixes'' the spin state of another dot. The presence of such non-local correlations breaks the assumption of bipartite dynamics. However, since every change of state is related to the act of electron tunneling to or from a single dot, one can still separate the dissipator $\mathcal{D}$ into the contributions $\mathcal{D}_1$ and $\mathcal{D}_2$ in a physically meaningful way [cf. Eqs.~\eqref{linddem} and~\eqref{dissloc}]. This enables us to describe the information flow within our approach.

\subsection{Equivalence with the classical approach for bipartite systems}
In the previous section we have shown that our approach can be applied to systems with a non-bipartite dynamics, for which the method proposed in Ref.~\cite{horowitz2014} is not applicable. Here, on the other hand, we demonstrate that when the dynamics is bipartite both approaches becomes equivalent; this can take place when the Hamiltonians of both subsystems commute with the total Hamiltonian. Therefore, our method can be seen as a direct generalization of the one proposed in Ref.~\cite{horowitz2014}.

Let us consider a two-component open quantum system consisting of subsystems $X$ and $Y$, with dynamics described by an effectively classical secular master equation. Furthermore, let us also assume that the Hamiltonian of the total system
\begin{align}
\hat{H}_S=\hat{H}_X+\hat{H}_Y+\hat{H}_{XY},
\end{align}
commutes with the Hamiltonians of the subsystems $\hat{H}_X$ and $\hat{H}_Y$. Then the eigenstates of the Hamiltonian $\hat{H}_S$ are products of the eigenstates of the Hamiltonians of subsystems: $|i \rangle=|x \rangle |y \rangle$, where $|i \rangle$ is an eigenstate of the total system whereas $|x \rangle$  ($|y \rangle$) is an eigenstate of the subsystem $X$ ($Y$). The state of the system can be expressed by the diagonal density matrix
\begin{align}
\rho=\sum_{x,y} p(x,y) |x \rangle |y \rangle \langle y| \langle x|,
\end{align}
where $p(x,y)$ is the probability of the state $|x \rangle |y \rangle$. The reduced density matrix of the subsystems $X$ and $Y$ will analogously read as
\begin{align}
\rho_X &=  \sum_x p(x) |x \rangle \langle x|, \\
\rho_Y &=  \sum_y p(y) |y \rangle \langle y|,
\end{align}
where
\begin{align}
p(x) &=\sum_y p(x,y), \\
p(y) &=\sum_x p(x,y),
\end{align}
are the probabilities of the states $|x \rangle$ and $|y \rangle$, respectively.

The dynamics of the state probabilities generated by the dissipator $\mathcal{D}$ is given by the rate equation (Eq.~(1) from Ref.~\cite{horowitz2014})
\begin{align}
d_t p(x,y)=\sum_{x',y'} \left[W_{x,x'}^{y,y'} p(x',y') - W_{x',x}^{y',y} p(x,y)\right],
\end{align}
where $W_{x,x'}^{y,y'}$ is the transition rate of the jump $(x',y') \rightarrow (x,y)$. 

Let us now assume that the dynamics is bipartite, i.e., that there is no jumps simultaneously changing states of the subsystems $X$ and $Y$. Mathematically this reads (Eq.~(2) from Ref.~\cite{horowitz2014})
\begin{equation}
W_{x,x^\prime}^{y,y^\prime}=
\left\{\begin{array}{cc}w_{x,x^\prime}^y & x\neq x^\prime; y=y^\prime \\
w_{x}^{y,y^\prime} & x= x^\prime; y\neq y^\prime \\
0 & {\rm otherwise}
\end{array}\right..
\end{equation}
Here the rates $w_{x,x^\prime}^y$ and $w_{x}^{y,y^\prime}$ correspond to the operation of the dissipator $\mathcal{D}_X$ and $\mathcal{D}_Y$, respectively. Dynamics of the subsystem $X$ is then described by the rate equation
\begin{align}
d_t p(x)=\sum_{x',y} \left[w_{x,x'}^{y} p(x',y) - w_{x',x}^{y} p(x,y)\right].
\end{align}
An analogous equation can be written for the subsystem $Y$. 

Following Eq.~(17) from the main text, the information rate associated with the subsystem $X$ can be calculated as
\begin{align}
\dot{I}_X =&-\text{Tr} \left( d_t \rho_X\ln \rho_X \right)+\text{Tr} \left[ \left( \mathcal{D}_X \rho \right) \ln \rho \right] \\ \nonumber
=&-\sum_{x} d_t p(x) \ln p(x) \\ \nonumber &+\sum_{x,x',y} \left[w_{x,x'}^{y} p(x',y) - w_{x',x}^{y} p(x,y)\right] \ln p(x,y) \\ \nonumber
=&\sum_{x,x',y} \left[w_{x,x'}^{y} p(x',y) - w_{x',x}^{y} p(x,y)\right] \ln \frac{p(x,y)}{p(x)} \\ \nonumber
=&\sum_{x \geq x',y} J_{x,x'}^{y} \ln \frac{p(x,y) p(x')}{p(x) p(x',y)}=\sum_{x \geq x',y} J_{x,x'}^{y} \ln \frac{p(y|x)}{p(y|x')}.
\end{align}
Here we have denoted $w_{x,x'}^{y} p(x',y) - w_{x',x}^{y} p(x,y)=J_{x,x'}^{y}$ and used the definition of the conditional probability $p(y|x)=p(y,x)/p(x)$. The result is equivalent to Eq.~(10) from Ref.~\cite{horowitz2014}. Upon putting on the obtained $\dot{I}_X$ into Eq.~(18) from the main text one obtains Eq.~(11) from Ref.~\cite{horowitz2014}. This demonstrates the equivalence of both approaches for systems with the bipartite dynamics.

\subsection{The unitary and the dissipative contribution to the information flow}
As briefly mentioned in the main text, the information flow $\dot{I}_i$ can be decomposed into two separate contributions associated with the unitary and the dissipative evolution of the system. In order to calculate them, let us first define the unitary and the dissipative contributions to the total derivative of the reduced density matrix $\rho_i$:
\begin{align}
\dot{\rho}_i^U &= -i\text{Tr}_j \left( \left[\hat{H}_S, \rho \right] \right), \\
\dot{\rho}_i^D &= \text{Tr}_j \left( \mathcal{D} \rho \right),
\end{align}
such that $d_t \rho_i = \dot{\rho}_i^U+\dot{\rho}_i^D$; here $\text{Tr}_j$ is the trace over the subsystem $j \neq i$. Following Eq.~(17) from the main text, the unitary and the dissipative contribution to the information flow can be calculated as
\begin{align}
\dot{I}_i^U =&-\text{Tr} \left( \dot{\rho}_i^U \ln \rho_i \right), \\
\dot{I}_i^D =&-\text{Tr} \left( \dot{\rho}_i^D \ln \rho_i \right)+\text{Tr} \left[ \left( \mathcal{D}_i \rho \right) \ln \rho \right],
\end{align}
such that $\dot{I}_i = \dot{I}_i^U+\dot{I}_i^D$.We emphasize that the separation into the unitary and the dissipative contribution is unambiguous, i.e., does not depend on the choice of the basis, which follows from the invariance of the trace under the change of basis~\cite{nielsen2010}.

%
\begin{figure}
	\centering
	\includegraphics[width=0.9\linewidth]{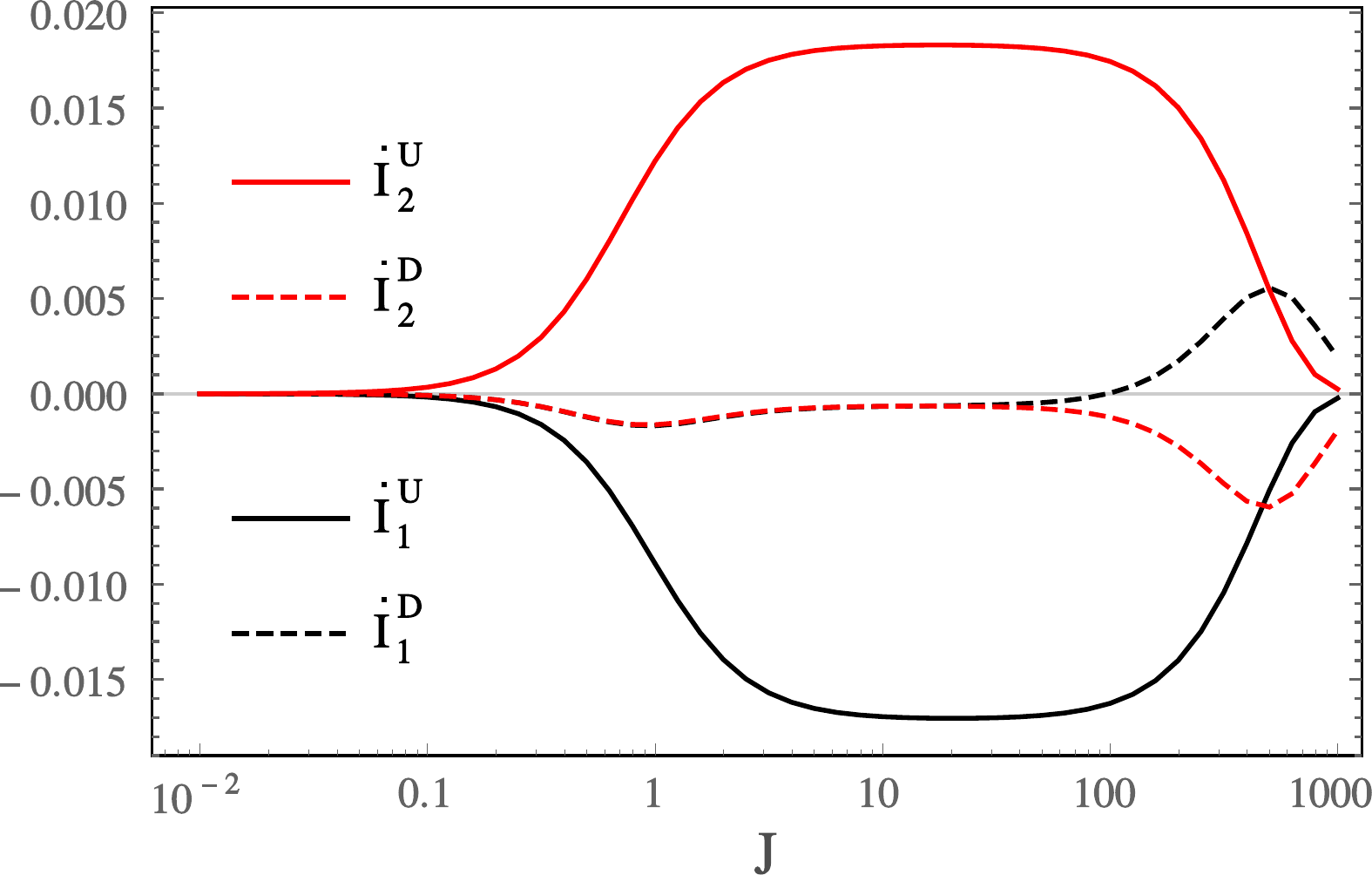}		
	\caption{The unitary and the dissipative contribution to the information flows $\dot{I}_1$ and $\dot{I}_2$ calculated using the non-secular Lindblad equation for parameters as in Fig.~\ref{fig:compsec}.}
	\label{fig:contr}
\end{figure}
%

Results for the autonomous Maxwell demon are presented in Fig.~\ref{fig:contr}. As one can see, in nearly whole range of $J$ the information flow is dominated by the unitary contribution associated with the coherent oscillations between the spin states. For small $J$ the dissipative contribution in both subsystems is negative. This is because the dissipative dynamics of both dots is effectively uncorrelated, and the uncorrelated dynamics can only decrease the mutual information (this is referred to as the data processing inequality~\cite{wilde2013}). Dissipation becomes correlated only for high $J \approx 500$. In this regime, unitary and dissipative contributions to the flow $\dot{I}_i$ have opposite signs, i.e., the dissipative dynamics suppress the information flow resulting from the coherent oscillations. Due to this fact, for high $J$ the system ceases to work as a Maxwell demon.

Interestingly, the unitary contribution to the information flow dominates even in the regime in which the secular approximation, which decouples populations and coherences, converges to non-secular approaches when it comes to the total information flow. However, by means of the secular master equation one cannot distinguish between different contributions to the information flow. This is another illustration of the fact that, although the secular approximation can describe the operation of the system as an autonomous Maxwell demon, it cannot fully account for its quantum coherent nature.